\documentclass{PoS}

\title{Charm Decays and Spectroscopy at BABAR}

\ShortTitle{Charm Decays and Spectroscopy at BABAR}

\author{\speaker{Romulus Godang}
\thanks{This work was supported by the U.S. Department of Energy under 
grant No. DE-FG02-96ER-40970}\\
        On Behalf of the BABAR Collaboration\\

\vspace{0.2cm}

        Department of Physics\\	
        University of South Alabama\\
        ILB 115, 307 University Blvd., N.\\
        E-mail: \email{godang@usouthal.edu}\\

\vspace{0.1cm}
SLAC-PUB-15329

}

\abstract{
We present searches for rare charm decays of the form $X_c^+ \to h^{\pm} \ell^{\mp} 
\ell^{(')^+}$, where $X_c^+$ is a charm hadron either $D^+$, $D_s^+$, or $\Lambda_c^+$, 
and $\ell^{(')^{\pm}}$ is an electron or muon. These modes are based on 
384 $fb^{-1}$ of $e^+ e^-$ annihilation data collected at the $\Upsilon(4S)$ 
resonance with the BABAR detector at the SLAC National Accelerator Laboratory.
We also present the flavor-changing neutral-current decays $D^0 \to e^+e^-$, 
$D^0 \to \mu^+ \mu^-$, and $D^0 \to e^{\pm} \mu^{\mp}$ that corresponds to 
an integrated luminosity of 468 $fb^{-1}$ of data.
The decay $D^0 \to e^+ \mu^-$ is further lepton-flavor violating,
and thus occur only through very slow neutrino mixing.
These decays constitute sensitive probes for possible new-physics
contribution. We report new limits on the branching fractions of
these decays. 
}

\FullConference{36th International Conference on High Energy Physics,\\
		July 4-11, 2012\\
		Melbourne, Australia}

\begin{document}

\section{INTRODUCTION}
In the Standard Model (SM), the flavor-changing neutral processes are very rare
and are of obvious interest in the search for new physics. In the Flavor-Changing Neutral 
Current (FCNC) decays $D^0 \to \ell^+ \ell^-$, where $\ell$ is an either electron or muon, 
are strongly suppressed by the Glashow-Iliopoulos-Maiani (GIM) mechanism~\cite{GIM}. 
These decays cannot occur at tree level in the SM. The branching fraction of the decays 
$D^0 \to \ell^+ \ell^-$ are predicted to be ${\cal{O}}(10^{-13})$~\cite{Burdman}. 

Most of the attention on FCNC decays has been focused in the $K$ and $B$ meson sectors and 
less in the charm meson sector. It is due to the fact that the SM expectations for 
$D^0-\bar{D^0}$ mixing are very small compared to the $K^0-\bar{K^0}$ and $B^0-\bar{B^0}$ mixing. 
However, the FCNC decay in the charm sector is unique due to its decays involve an up-type 
quark which implies into an effective GIM cancellations and new physics.
The decay modes of Lepton-Flavor Violating (LFV) which corresponding to two leptons with 
two oppositely charged of different flavor and Lepton-Number Violating (LNV) decays where
two leptons have the same charge are forbidden in the SM. 

\begin{figure}[!htb]
\begin{center}
\includegraphics[width=4.4cm]{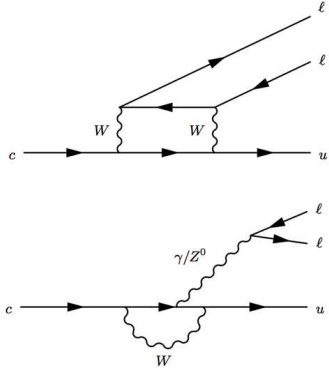}  
\caption{Standard Model short-distance contributions to the $c \to u \ell^+ \ell^-$ transition.}
\label{fig:c_to_uellell}
\end{center}
\end{figure}

Figure~\ref{fig:c_to_uellell} shows the Standard Model short-distance contributions to
the $c \to u \ell^+ \ell^-$ transition. The branching fraction for the decay of
$D \to X_u \ell^+ \ell^-$ is predicted to be ${\cal{O}}(10^{-8})$~\cite{Burdman, Fajfer2001}.   
The decay of $c \to u \ell^+ \ell^-$ is screened by the long distance contributions. It is also
expected to dominate over the short distance contributions in $D^0-\bar{D^0}$ mixing. 
The long distance contributions were shown to be largely dominant in $c \to u \ell^+ \ell^-$.
The experimental upper bounds on the branching fraction of $c \to u \ell^+ \ell^-$ is presently
in the range of ${\cal{O}}(10^{-5})$~\cite{aitala}. It is an order of magnitude larger than
the Standard Model prediction for specific channels~\cite{Fajfer1998}. 
The highest rate of $D \to V \ell^+ \ell^-$ channel with $V = \rho, \omega, \phi, K^*$ is
the decay of $D_s^+ \to \rho^+ \ell^+\ell^-$. It is predicted at the highest rate 
$\approx 3 \times 10^{-5}$, however the there are unfortunately no experimental
data on this channel.

\section{The BABAR DETECTOR AND DATA SET}
The BABAR detector was operated at the PEP-II asymmetric-energy storage rings at
the SLAC National Accelerator Laboratory. The data used in this analysis were 
collected with the BABAR detector. The data sample consist of an integrated
luminosity of 384 $fb^{-1}$ for $X_c^+ \to h^{\pm} \ell^{\mp} 
\ell^{(')^+}$ and 468 $fb^{-1}$ for $D^0 \to \ell^+ \ell^-$ accumulated at the $\Upsilon(4S)$ 
resonance and 40 MeV below the $\Upsilon(4S)$ resonance.
The asymmetric energy of the PEP-II $e^+$ and $e^-$ beams result in 
a Lorentz boost $\beta\gamma \approx 0.55$ of the $B\bar{B}$ pairs.

A detail description of the BABAR detector is presented elsewhere~\cite{Babar_nim}.
The momenta of the charged particles are measured in a tracking system consisting 
of a 5-layer double sided silicon vertex tracker (SVT) and a 40-layer drift
chamber (DCH). The SVT and DCH operate within a 1.5 T solenoid field and have a combined 
solid angle coverage in the center of mass frame of 90.5\%. A detector of internally reflected
Cerenkov radiation (DIRC) is used for charged particle identifications of pions, kaons, 
and protons with likelihood ratios calculated from $dE/dx$ measurements in the SVT and DCH.  
Photons and long-lived neutral hadrons are detected and their energies are measured in 
a CsI(Tl) electromagnetic calorimeter (EMC). For electrons, energy lost due to 
bremsstrahlung is recovered from deposits in the EMC.

\section{ANALYSIS}
We select charm hadron candidates $X_c$ with center of mass frame momentum greater than
2.5 GeV/c to suppress combinatoric background. The $e^+ e^-$ invariant mass is required 
to be greater than 200 Mev/$c^2$ in order to reject photon conversion and $\pi^0$ decays
to $e^+ e^- \gamma$. For the $D_{(s)}^+ \to \pi \phi$, $\phi \to \ell^+ \ell^-$ decay mode,
we excluded events with $0.95 < m(e^+ e^-) < 1.05$ GeV/$c^2$ and  
$0.99 < m(\mu^+ \mu^-) < 1.05$ GeV/$c^2$ to reject the decays through the $\phi$ resonance. 
The QED backgrounds was suppressed by requiring at least five tracks in the event and 
that hadron candidate be consistent with the electron hypothesis.
After the initial event selection, significant combinatorial background contribution, 
we use three discriminating variables in likelihood ratio: charm hadron candidate, 
total reconstructed energy in the event, and flight length significance.  

\begin{figure*}[tb]
\centerline{\includegraphics[width=10cm]{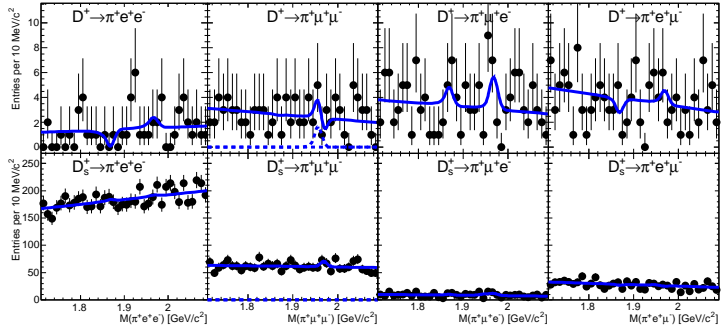}}
\caption{Invariant-mass distributions for
$D^+ \to \pi^+ \ell^+ \ell^{(\prime)-}$ (top) and $D_s^+ \to \pi^+ \ell^+ \ell^{(\prime)-}$ 
(bottom) candidates. The fit results shown in the solid line.}
\label{fig:Dppill} 
\end{figure*}

\begin{figure*}[tb]
\centerline{\includegraphics[width=10cm]{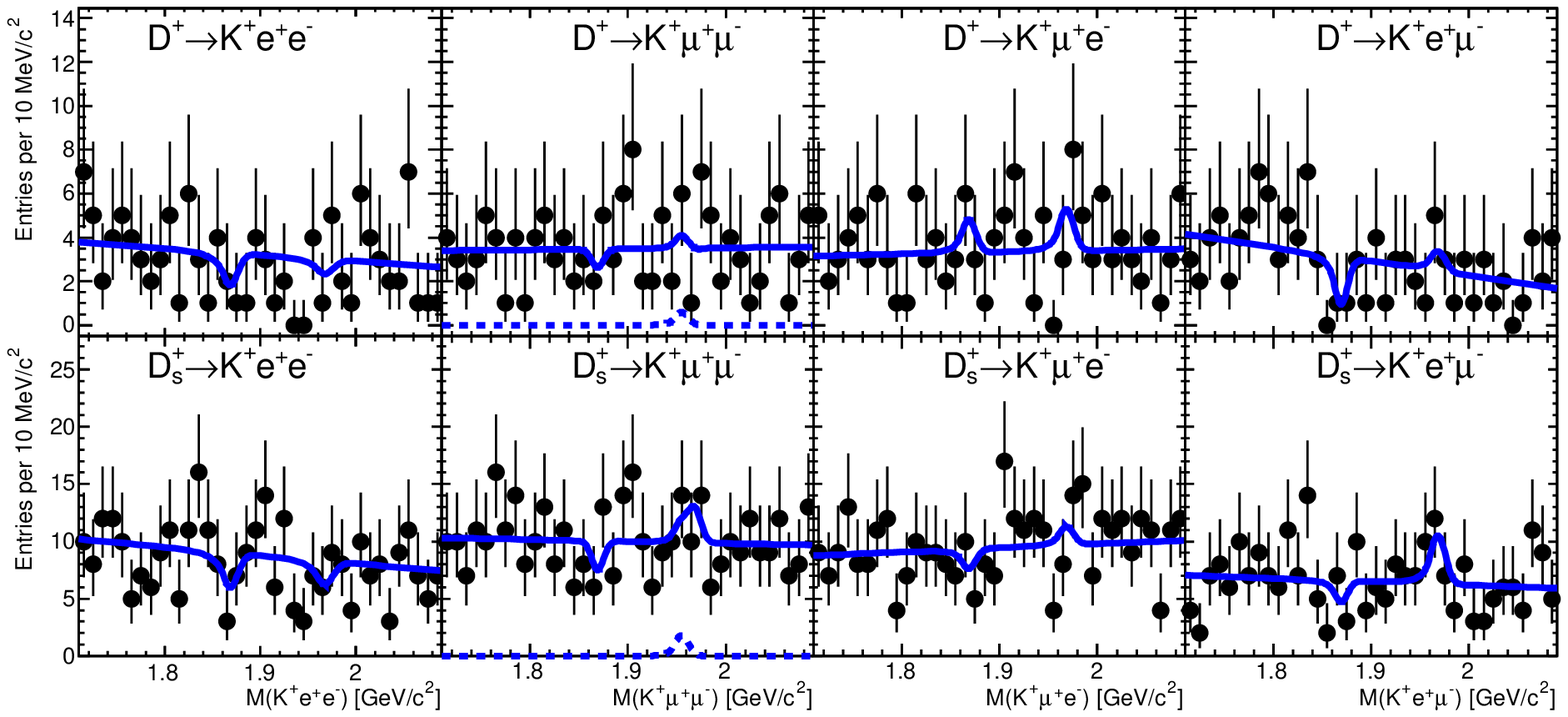}}
\caption{Invariant-mass distributions for
$D^+ \to K^+ \ell^+ \ell^{(\prime)-}$ (top) and $D_s^+ \to K^+ \ell^+ \ell^{(\prime)-}$ 
(bottom) candidates. The fit results shown in the solid line.}
\label{fig:DpKll}
\end{figure*}

\begin{figure*}[tb]
\centerline{\includegraphics[width=10cm]{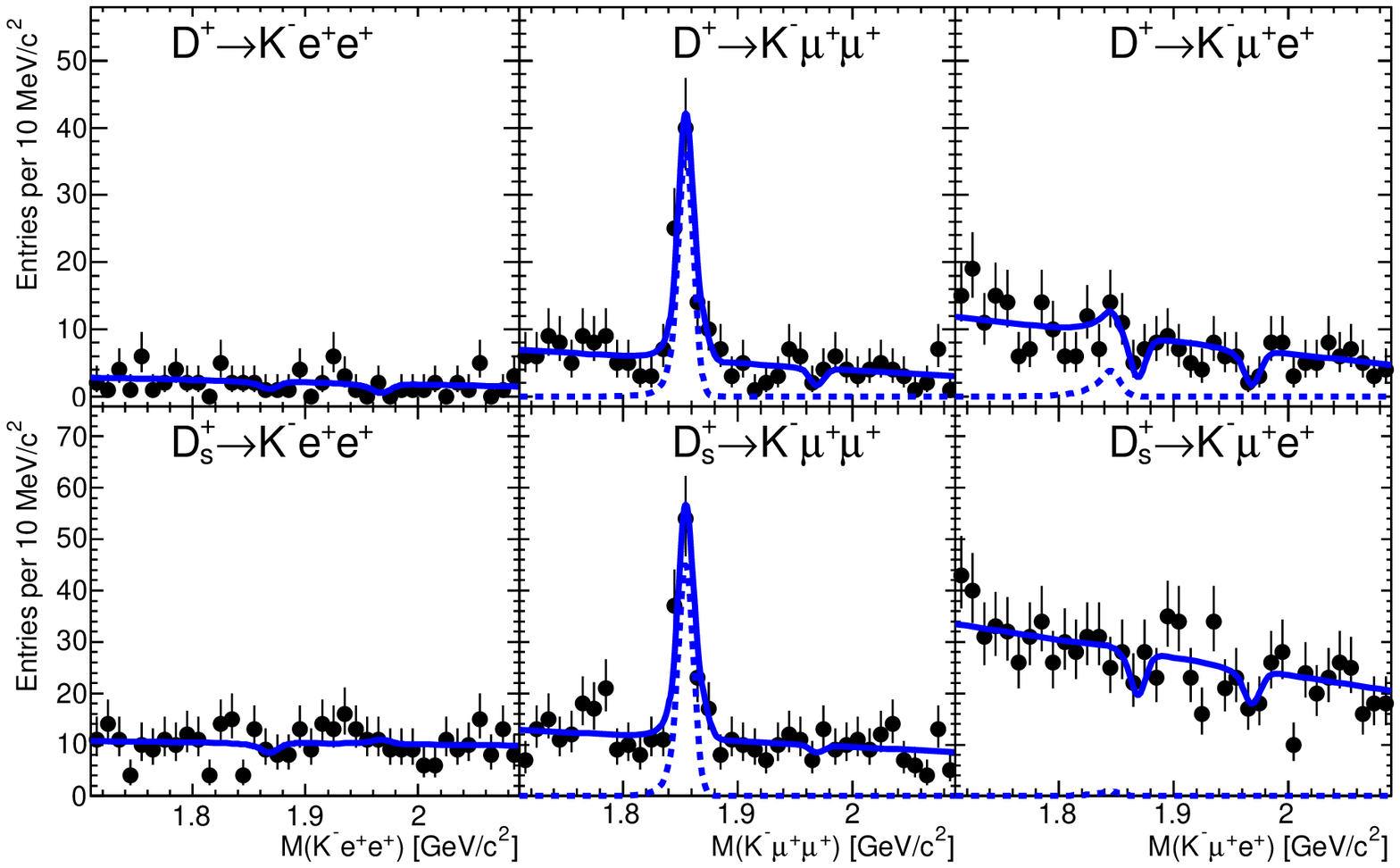}}
\caption{Invariant-mass distributions for $D^+ \to K^- \ell^+ \ell^{(\prime)+}$ (top)
and $D_s^+ \to K^- \ell^+ \ell^{(\prime)+}$ (bottom) candidates. 
The fit results shown in the solid line.}
\label{fig:DpKllLN}
\end{figure*}

\begin{figure*}[tb]
\centerline{\includegraphics[width=12cm]{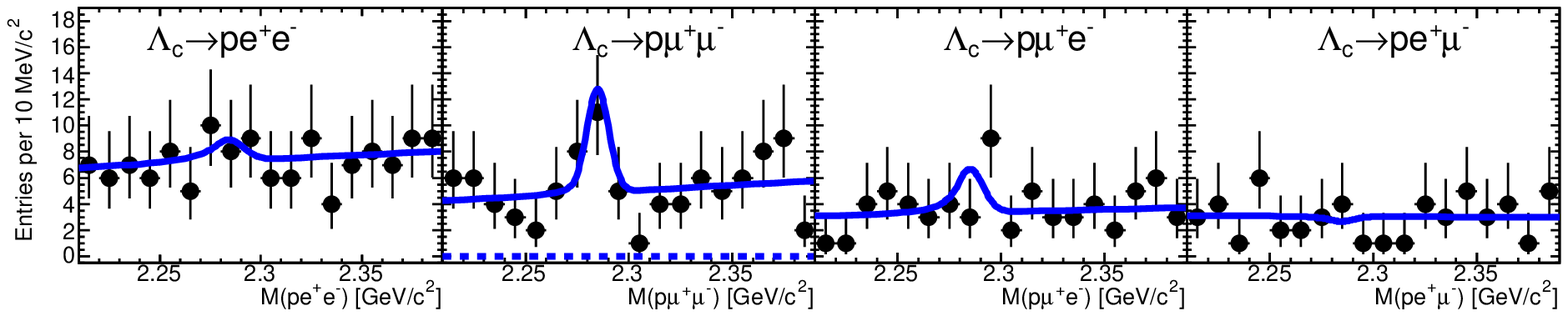}}
\caption{Invariant-mass distributions for $\Lambda_c^+ \to p \ell^+ \ell^{(\prime)-}$ 
candidates. The fit results shown in the solid line.}
\label{fig:Lcpll}
\end{figure*}

\begin{figure*}[tb]
\centerline{\includegraphics[width=12cm]{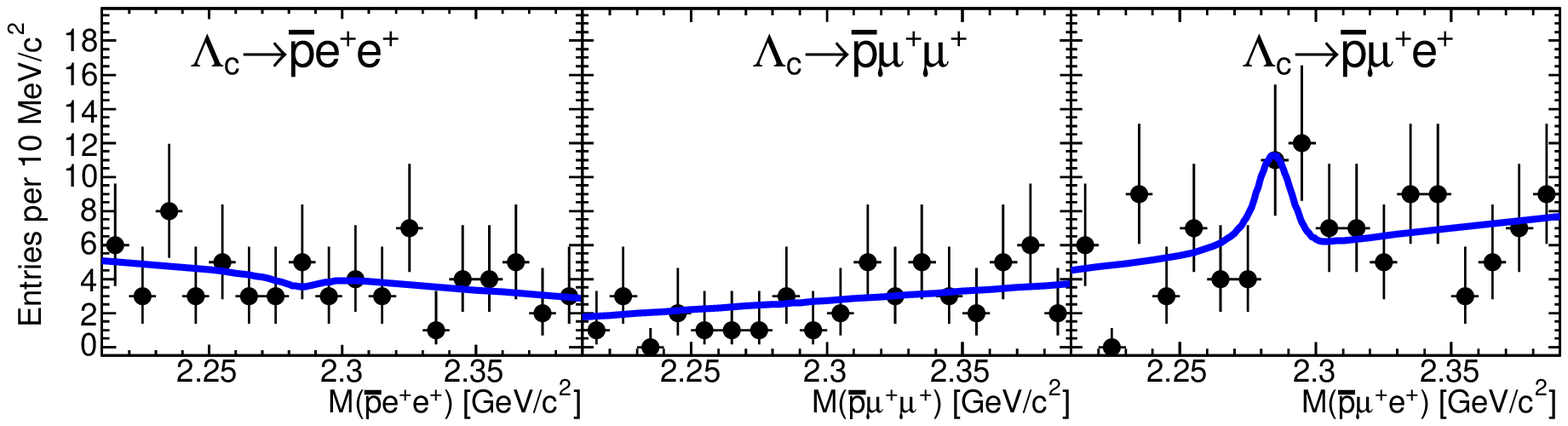}}
\caption{Invariant-mass distributions for $\Lambda_c^+ \to \bar{p} \ell^+ \ell^{(\prime)+}$
candidates. The fit results shown in the solid line.}
\label{fig:LcpllLN}
\end{figure*}

To measure the signal events we use extended, unbinned, maximum-likelihood.
These signals are converted to the known charm branching fractions by normalization. 
To reduce the systematic effects we choose normalization modes with kinematics similar to 
the kinematic of the signal decays. For decays of $D^+$ and $D_s^+$ mesons, 
the normalization mode is $\pi^+\phi$ where $\phi \to K^+K^-$. For the decays of 
$\Lambda_c^+$, we choose the decays of $\Lambda_c^+ \to p K^- \pi^+$. 
Figures~\ref{fig:Dppill}--\ref{fig:LcpllLN} show the fitting results of the invariant mass
of $X_c^+ \to h^{\pm} \ell^{\mp} \ell^{(')^+}$ decays.
The dashed curves show the background components for the dimuon modes in which muon 
candidates arise from misidentified hadrons.
Detail information on the likelihood selection, fitting procedure, systematic uncertainties, 
and fit results are available here~\cite{publication1}. 
We calculate the upper limits on the ratio of the branching factions at 90\% confidence level
(CL): 
${\cal{B}}(D_{(s)}^+ \to \pi^{\pm}\ell^{\mp}\ell^{(')+})/
{\cal{B}}(D_{(s)}^+ \to \pi^+\phi)$, 
${\cal{B}}(D_{(s)}^+ \to K^{\pm}\ell^{\mp}\ell^{(')^+})/
{\cal{B}}(D_{(s)}^+ \to \pi^+\phi)$, and 
${\cal{B}}(\Lambda_c^+ \to p^{(-)}\ell^{\mp}\ell^{(')^+})/
{\cal{B}}(\Lambda_c^+ \to p K^-\pi^+)$. The most significant signal is seen in the
decay of $\Lambda_c^+ \to p \mu^+\mu^-$ with yield of $11.1 \pm 5.0(stat)\pm2.5(syst)$.
It has a statistical significant of $2.6\sigma$. It is corresponding to 90\% CL upper
limit on the branching fraction of $44 \times 10^{-6}$.

\begin{figure*}
\begin{center}
\includegraphics[width=0.31\linewidth]{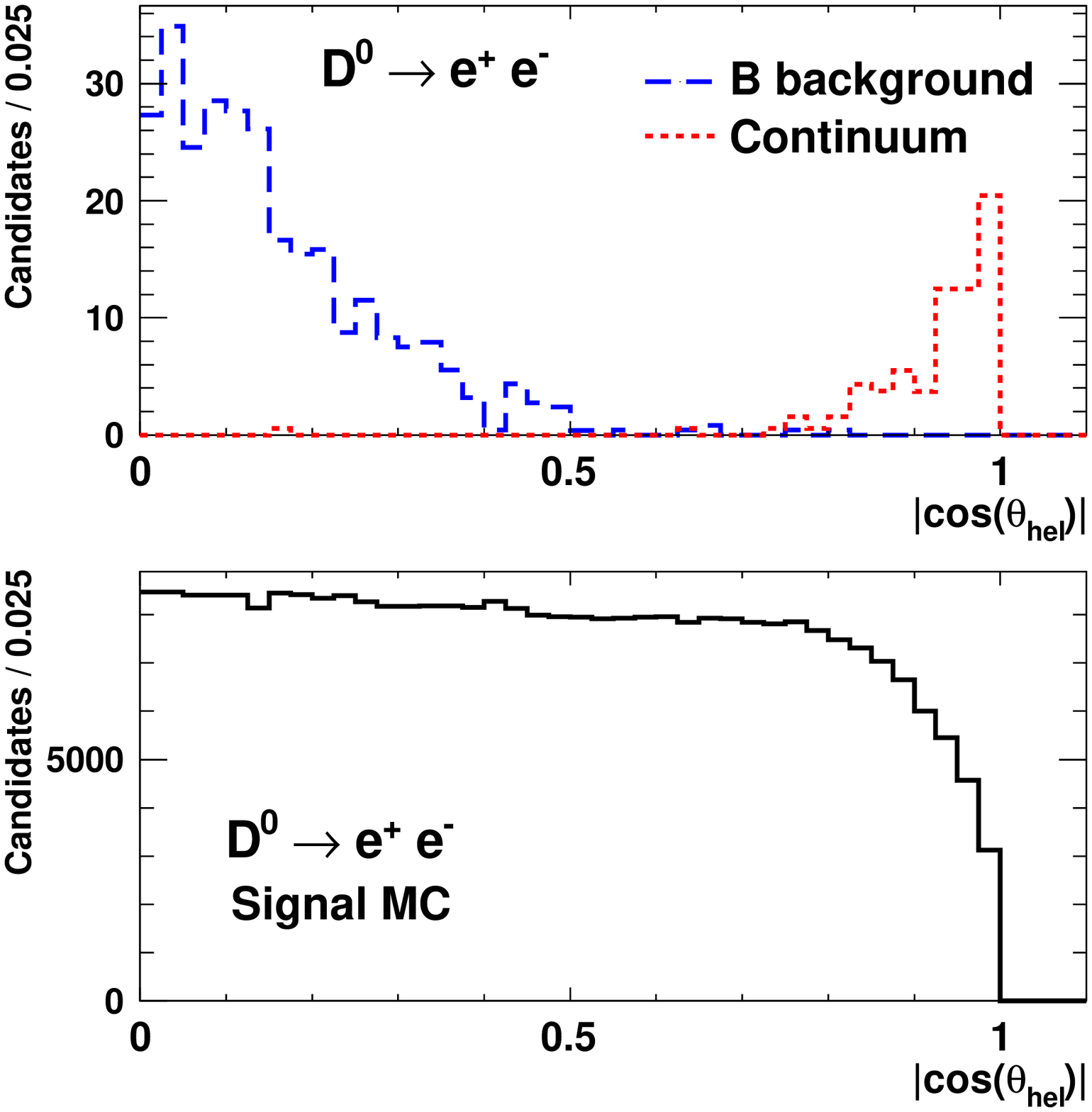}
\includegraphics[width=0.31\linewidth]{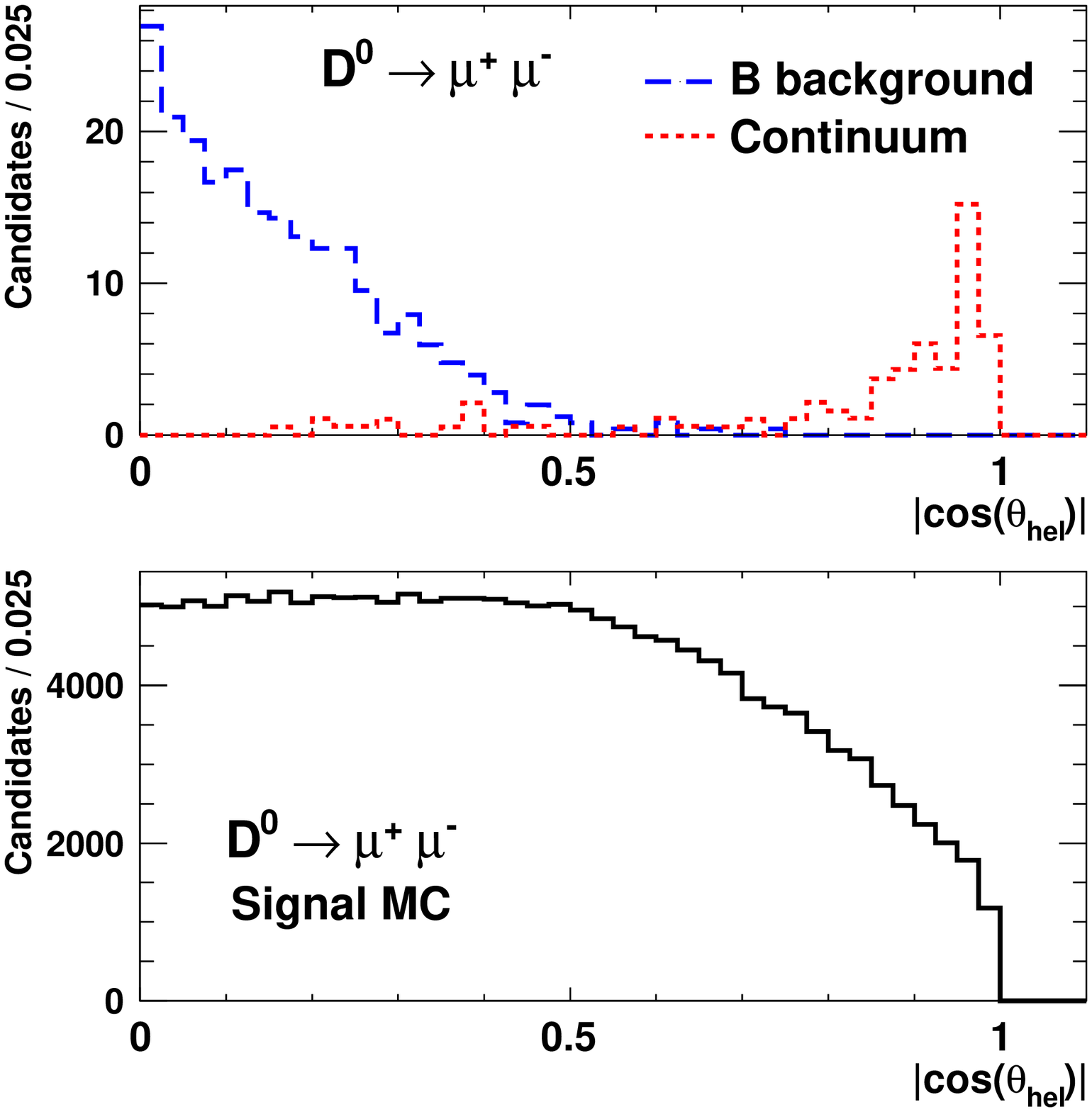}
\includegraphics[width=0.31\linewidth]{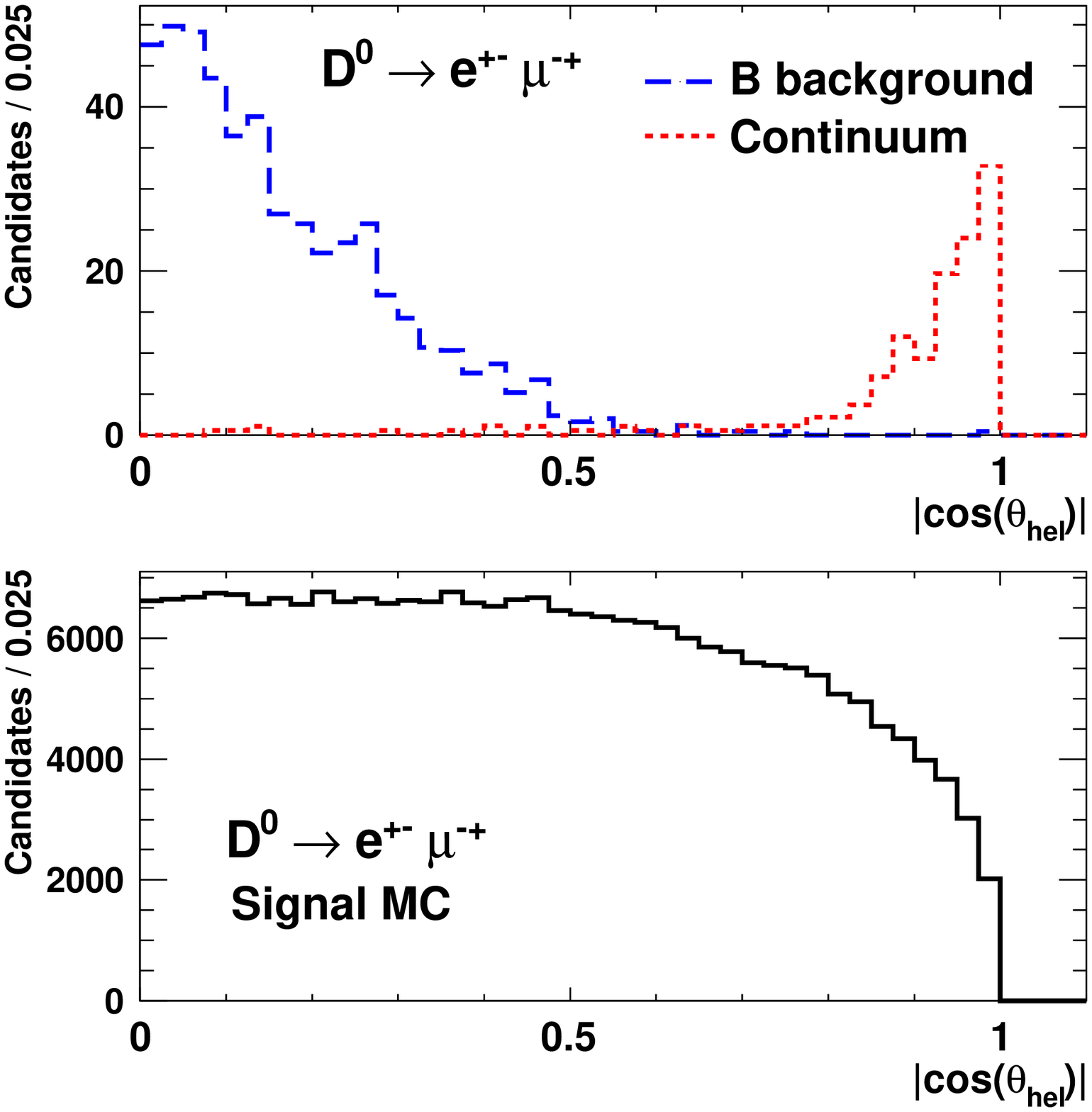}
\caption{Distributions of $|\cos(\theta_{\rm hel})|$ for the three signal channels:
$D^0 \to e^+ e^-$ (left), $D^0 \to \mu^+ \mu^-$ (center), and $D^0 \to e^{\pm}\mu^{\mp}$ 
(right). The bottom distributions show the signal Monte Carlo samples with an arbitrary normalization.}
\label{fig:coshel}
\end{center}
\end{figure*}

We also recently measured the flavor-changing neutral-current decays $D^0 \to e^+e^-$, 
$D^0 \to \mu^+ \mu^-$, and $D^0 \to e^{\pm} \mu^{\mp}$ that corresponds to an integrated 
luminosity of 468 $fb^{-1}$ of data. To normalize the decays of $D^0 \to \ell^+ \ell^-$,
we use $D^0 \to \pi^+\pi^-$ control sample and applying a linear combination
of Fisher discriminant~\cite{fisher} of the following five variables:
measured $D^0$ flight length, $|cos \theta_{\rm hel}|$ angle between the momentum of 
the positively-charged $D^0$ daughter and the boost direction from the lab frame 
to the $D^0$ rest frame (all in the $D^0$ rest frame), the missing transverse momentum 
with respect to the beam axis, the ratio of the $2^{\rm nd}$ and $0^{\rm th}$ 
Fox-Wolfram moments~\cite{fox-walfram}, and the $D^0$ momentum in the center of mass frame.
To remove the continuum combinatoric background we use the $|cos \theta_{\rm hel}|$ variable. 
Figure~\ref{fig:coshel} shows distributions of $|cos \theta_{\rm hel}|$ before applying 
a minimum Fisher discriminant. 

The branching fraction of $D^0 \to \ell^+\ell^-$ is given by the following expressions:
\begin{equation}
{\cal B}_{\ell\ell} = \left(\frac{N_{\ell\ell}}{N_{\pi\pi}^{\rm fit}}\right)
\ \left(\frac{\epsilon_{\pi\pi}}{\epsilon_{\ell\ell}}\right)
\ {\cal B}_{\pi\pi}
\ = \ S_{\ell\ell} \ \cdot \ N_{\ell\ell}
\end{equation}
where $S_{\ell\ell}$ is defined by
\begin{equation}
S_{\ell\ell} \ \equiv \ \frac{ {\cal B}_{\pi\pi} }
{    N_{\pi\pi}^{\rm fit} }
\frac{ \epsilon_{\pi\pi} }{ \epsilon_{\ell\ell} }.
\end{equation}
and $N_{\rm obs}$ is defined by
\begin{equation}
      N_{\rm obs} \ = \ {\cal B}_{\ell\ell} / S_{\ell\ell} + N_{BG}.
\end{equation}
The $N_{\ell\ell}$ and $N_{\pi\pi}^{\rm fit}$ are the number of $D^0 \to \ell^+\ell^-$ and
$D^0 \to \pi^+ \pi^-$ candidates, respectively. The ${\cal B}_{\pi\pi} = (1.400 \pm 0.026) 
\times 10^{-3}$~\cite{PDG2010}. We use the likelihood ratio ordering principle of Feldman and 
Cousins~\cite{feldman} to determine 90\% CL intervals.
We find one event of $D^0 \to e^+ e^-$ with background of $1.0 \pm 0.5$ events and 
two events of $D^0 \to e^{\pm} \mu^{\mp}$ with background of $1.4 \pm 0.3$ events.
These correspond to the 90\% CL upper limits for the branching fractions
$<1.7\times 10^{-7}$ for $D^0 \to e^+ e^-$ and $<3.3 \times 10^{-7}$ 
for $D^0 \to e^{\pm} \mu^{\mp}$. For the $D^0 \to \mu^+ \mu^-$ channel, we find
eight events with expected background of $3.9 \pm 0.6$. This corresponds to
90\% CL upper limits on the branching fraction of $[0.6, 8.1]\times 10^{-7}$.
Detail information on the likelihood selection, fitting procedure, systematic uncertainties, 
and fit results are available here~\cite{publication2}.

\section{CONCLUSIONS}
We have searched for the decay modes $D_{(s)}^+ \to \pi^{\pm}\ell^{\mp}\ell^{(')+}$, 
$D_{(s)}^+ \to K^{\pm}\ell^{\mp}\ell^{(')^+}$, and $\Lambda_c^+ \to p^{(-)}\ell^{\mp}
\ell^{(')+}$. No signals are observed and we report upper limits on 35 different branching 
ratios between $0.4\times10^{-4}$ and $37\times10^{-4}$ at 90\% CL. This corresponds to limits
on the branching fractions between $1\times 10^{-6}$ and $44\times
10^{-6}$.  

We also have searched for the leptonic charm decays $D^0 \to e^+ e^-$, $D^0 \to \mu^+ \mu^-$,
and $D^0 \to e^{\pm} \mu^{\mp}$. We find no statistically significant excess over the 
expected background. These results supersede our previous results~\cite{previousbabar} 
and are consistent with the results of the Belle experiment~\cite{belle}.

\section{ACKNOWLEDGMENTS}
The author would like to thank the organizers of ICHEP 2012, the $36^{th}$
International Conference on High Energy Physics, Melbourne, Australia. The supports from
the BABAR Collaboration, the University of South Alabama, and the University of Mississippi 
are gratefully acknowledged. 


\end{document}